\documentclass[preprint]{elsart}

\usepackage{natbib}
\usepackage{float}
\usepackage{subfigure}
\usepackage{rotating}
\usepackage{longtable}
\usepackage{amsmath}
\usepackage{setspace}

\usepackage{graphicx}

\usepackage{amssymb}

\usepackage{lineno}

\journal{Icarus}

\begin{document}

\footnotesize

\begin{frontmatter}

\title{Families among high-inclination asteroids}

\author[bg1]{Bojan Novakovi\'c}
\ead{bojan@matf.bg.ac.rs}
\author[to]{Alberto Cellino}
\author[bg2]{Zoran Kne\v zevi\'c}

\address[bg1]{Department of Astronomy, Faculty of Mathematics, University of Belgrade,
Studentski trg 16, 11000 Belgrade, Serbia}
\address[to]{INAF--Osservatorio Astronomico di Torino, Via Osservatorio 20, I-10025
Pino Torinese, Italy}
\address[bg2]{Astronomical Observatory, Volgina 7,
         11060 Belgrade 38, Serbia}

\begin{abstract}
We present a new classification of families identified among the
population of high-inclination asteroids. We computed synthetic
proper elements for a sample of 18,560 numbered and multi-opposition
objects having sine of proper inclination greater than $0.295$. We
considered three zones at different heliocentric distances (inner,
intermediate and outer region) and used the standard approach based on the Hierarchical
Clustering Method (HCM) to identify families in each zone. In
doing so, we used slightly different approach with respect to previously
published methodologies, to achieve a more reliable and robust
classification. We also
used available SDSS color data to improve membership and identify
likely family interlopers. We found a total of 38 \textit{families}, as well as
a significant number of \textit{clumps} and \textit{clusters} deserving further
investigation.
\end{abstract}


\end{frontmatter}

\section{Introduction}
\label{s:intro}

As first realized by \citet{hirayama1918}, some concentrations of
asteroids are apparent if we look at their distribution in the space
of orbital elements. These groups, known as the asteroid families,
are believed to have originated from catastrophic disruptions of
single parent bodies as a consequence of energetic asteroid
collisions. These events are thought to have produced ejections of
fragments into nearby heliocentric orbits, with relative velocities
much lower than the parent body's orbital speed. Asteroid families
were extensively investigated in the last decades, because these are
unique \textit{natural laboratories} to study the outcomes of
high-energy collisions \citep{zapetal2002,michel03,durda07}. Also, the
number of currently identified families is an important constraint
to model the collisional history of the asteroid main belt \citep{bottke05}.

Asteroid families are usually identified in the space of proper
elements: proper semi-major axis ($a_{p}$), proper eccentricity
($e_{p}$), and proper inclination ($I_{p}$). Proper orbital
elements, being quasi-integrals of motion and thus nearly constant
over time, are suited to be used to identify groupings that are
stable and are not affected by transient oscillations of
the osculating orbital elements.

To date, several tens of families have been discovered across the
asteroid main belt \citep[e.g.][]{zappala95,bend02,nes2005}. Most of
these families are located at proper inclinations lower than about
$17^{\circ}$ ($\sin(I_{p}) \leq 0.3$). The situation is more difficult
at higher inclinations. The number of existing asteroids tends to
decrease for increasing orbital inclination. As a consequence,
asteroid surveys are usually centered around the ecliptic, and this
also tends to introduce an observational bias against the discovery
of high-inclination objects. Until recently, the number of known
high-inclination asteroids was relatively small. Moreover, the
number of high-inclination asteroids for which proper elements had
been computed was even smaller. This was due to the fact that
analytical proper elements \citep{MilKne90,MilKne94}, which are
computed for both numbered and multi-opposition asteroids,
are not accurate enough for highly inclined orbits. In the past,
computations of proper elements by means of the methods specially developed to handle highly
inclined and/or eccentric orbits were carried out by some authors
for a limited number of asteroids
\citep{lemaitre1994}. More recently, the computation of the
so-called synthetic proper elements \citep{synthpro1,synthpro2}, has
made it possible to compute with a good accuracy proper elements for
high-inclination and high-eccentricity orbits as well. A large data
set of asteroid proper elements is essential for the identification
of asteroid families. Previous searches for families among
high-inclination asteroids were seriously limited by the paucity of
discovered asteroids and available proper elements. This problem affected also
the only systematic search published in the recent years, namely that
performed by \citet{gil_hutton2006}.

\begin{figure}[ht]
\centering
\includegraphics[height=10.0cm,angle=-90]{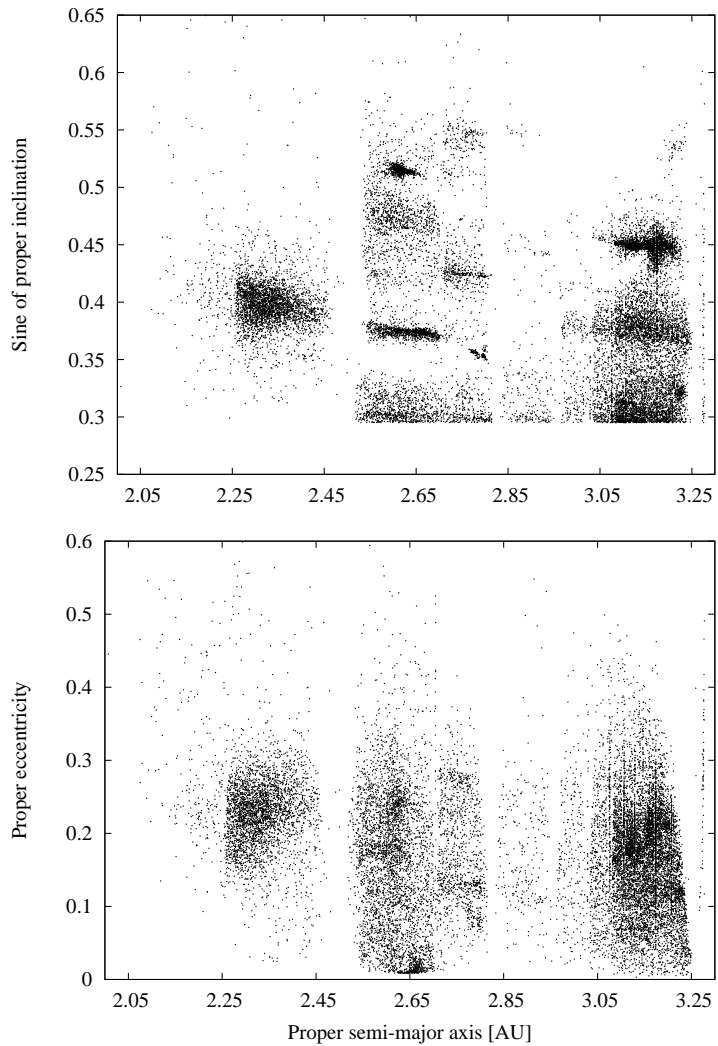}
\caption[]{The distribution of 18,560 highly inclined asteroids
considered in the present analysis in the ($a_{p}$,$\sin(I_{p})$) (top)
and ($a_{p}$,$e_{p}$) (bottom) planes.}\label{f:aei}
\end{figure}

Recently, the number of known high-inclination
asteroids has increased significantly. As of June 2010, when we
commenced the present analysis, the database of synthetic proper
elements, maintained at \textit{AstDys} web
page\footnote{http://hamilton.dm.unipi.it/astdys/index.php?pc=5},
included 10,265 objects with $\sin(I_{p})$ greater than $0.295$.
In this database, however, synthetic
proper elements were available only for numbered asteroids. In order
to increase the available sample, following the approach described
in \citet{synthpro2}, we have computed synthetic proper elements for
an additional sample of 8295 multi-opposition objects. In this way,
for the purposes of our analysis, we have used a data-base of
synthetic proper elements including 18,560 objects.\footnote{The
accuracy of this proper element data set is similar to that of a
recently analyzed sample of Hungaria asteroids \citep{milani2010}. A
slightly lower accuracy overall in proper eccentricity is likely due
to the fact that our sample includes a larger number of highly
eccentric orbits. The overall quality of the proper elements at our
disposal is in any case fully appropriate for the purposes of our
analysis.} The distributions of these asteroids, in the
($a_{p}$,$e_{p}$) and ($a_{p}$,$\sin(I_{p})$) planes, are shown in
Fig.~\ref{f:aei}.
As a comparison, in his search for high-inclination families
\citet{gil_hutton2006} used a sample about 5 times smaller (3697
asteroids).\footnote{Instead of the limit of $\sin(I_{p})$ grater
than $0.3$, used by \citet{gil_hutton2006}, we chose to work with
objects having $\sin(I_{p})$ greater than $0.295$. This was done in
order to identify possible traces of the classical main belt families
among the highly inclined asteroids. In our sample
there are 17,564 asteroids with $\sin(I_{p})$ grater than $0.3$.}
Similar studies were also performed for the
inner and intermediate regions of the asteroid belt by
\citet[][]{carruba2009,carruba2010b}, using only 1736
objects in the inner region, and 4452 objects in the intermediate
zone. Our sample is about 2 times larger in the inner zone, and more than $20\%$ larger
in the intermediate zone (3553 and 5439 asteroids respectively).
No search for families in the outer belt was performed so far in the
investigations by Carruba.

The purpose of this paper is to present a systematic search for
families among high-inclination asteroids. The paper is organized as
follows. In Section~\ref{s:method} the methods and techniques used
to identify statistically significant groups of asteroids in the
space of proper elements are described. The results of this analysis
are presented in Section~\ref{s:results}. These include a list of
different kinds of groupings that we found in the inner,
intermediate and outer part of the asteroid belt. In
Section~\ref{s:sdss} we improve our analysis by taking into account
the information about the object colors as obtained by the Sloan
Digital Sky Survey (SDSS). Finally, in
Section~\ref{s:conclusions} we outline our main conclusions.

\section{The Family Identification Method}
\label{s:method}

In our analysis we have in general followed the approach described in
the papers by \citet{zappala90,zappala94,zappala95}. Here we
briefly summarize the main principles of this approach, we describe
the main steps of its practical implementation, and explain a few
modifications that we have introduced, mainly to improve
the reliability of family membership.

We identified asteroid families in our proper element data-base by
using the so-called Hierarchical Clustering Method (HCM) based upon
the nearest-neighbor concept \citep[see e.g.][]{zappala90}. In
general terms, the HCM approach is based on a few simple ideas.
First, a metric is defined, to compute mutual distances between the
objects in the space of proper elements. In particular, we have
adopted the same metric that was used in previous papers. Therefore,
the distance $d$ between two objects is computed according to the
relation:
\begin{equation}
d = n a_{p}\sqrt{\frac{5}{4}(\frac{\delta a_{p}}{a_{p}})^{2}+2(\delta e_{p})^{2}+2(\delta sin(I_{p}))^{2}}
\label{eq:metric}
\end{equation}
where $na_{p}$ is the heliocentric velocity of an asteroid on a
circular orbit having the semi-major axis $a_{p}$. $\delta
a_{p}$~=~$a_{p_{1}}-a_{p_{2}}$, $\delta
e_{p}$~=~$e_{p_{1}}-e_{p_{2}}$, and $\delta
\sin(I_{p})$~=~$\sin(I_{p_{1}})-\sin(I_{p_{2}})$. The indexes (1) and (2)
denote the two bodies under consideration. Note also that $a_{p}$ in
the above formula corresponds to the average of $a_{p_{1}}$ and
$a_{p_{2}}$.

With this choice, $d$ has the dimension of a velocity, and is
usually expressed in m/s. Once the mutual distances are computed, it
is possible to identify the existence of groupings formed by objects
that, at a given level of distance $d$, have distances from
their closest neighbor smaller than $d$. The so-called
\emph{stalactite diagrams}, first used by \citet{zappala90}, are an
effective way to display the groupings found at different distance
levels, and to show how the membership of each group varies as a
function of the distance limit. Examples will be given in the next
section. Using this representation, the groupings of objects present
in a given sample are graphically displayed as a system of
\textit{stalactites}, the most compact groupings being represented
as the deepest stalactite branches.

The basic problem with the HCM approach is to define criteria for groupings
that cannot be due to pure chance. The simple idea is that asteroid
families produced by collisional processes should show up as deep
and thick stalactite branches which cannot be produced by other
mechanisms. To put this in more quantitative terms, in the classical
papers adopting the HCM approach, a critical value of distance
$d_{c}$ was found, for which it could be reasonably concluded that
groupings giving rise at deeper stalactite branches, or stalactites
reaching the same distance level, but with unlikely high numbers of
members, could not be due to chance, and necessarily have a physical
origin.

In its practical implementation, the HCM approach includes therefore
two basic parameters which have to be defined. One is the cut-off
distance $d_{c}$. The second parameter is the minimum number of
objects, $N_{crit}$, that is requested to characterize a
statistically significant group (with respect to our
selection criteria) at $d_{c}$.

As for $d_{c}$, in the previous papers this parameter was derived by
creating artificial populations ("Quasi Random Populations") of
synthetic objects, equal in number to the real population present in
a given region of the proper element space, and built in such a way
as to mimic independently the large-scale distributions of $a_{p}$,
$e_{p}$, and $\sin(I_{p})$ of the real population. The distance
levels of the deepest stalactite branches formed by a minimum number
of $N_{crit}$ synthetic objects were recorded. By repeating this
operation several times, an average value of minimum distance at
which random grouping could still occur was derived, together with
its uncertainty. This distance level was called Quasi Random Level
(QRL), and was assumed to correspond to $d_{c}$ for a given region
of the proper element space. QRL built in such a way takes into
account non-homogeneity of the distribution of the real objects in 
the given region. Thus, we can derive an evaluation of the distance
level in that region for which we cannot expect that denser clusters
of objects could exist purely due to chance. This takes into account
the features of the non-homogeneous proper element distribution of
the real objects in the given region. This approach is fairly
powerful, but it is not totally exempt from problems. For instance,
the obtained QRL is an average value of distance for a given region,
and it ``smears'' out the small-scale, local properties of the
distribution of objects. The main undesired consequence of this
concerns the resulting definition of family memberships. The QRL
concept is very satisfactory for the identification of families in a
given region of the proper elements space, but it turns out to be a
little too rigid as far as family membership is concerned, because
the membership of a family tends to be more influenced by the local
density of objects. For this reason, as we will explain below, in
this paper we have modified the criteria adopted in previous
application of the HCM for defining family memberships.

As for the adopted value of $N_{crit}$, it was chosen to be $5$ at
the epoch of the analysis performed by \citet{zappala90}. In
subsequent analyzes considering increasingly bigger data-sets of
asteroid proper elements \citep{zappala94,zappala95}, the adopted
values of $N_{crit}$ were scaled as the square root of the ratio
between the numbers of objects in the newer and in the older sample.
In particular, the values of $N_{crit}$ used by \citet{zappala95}
were ($10$, $9$, $8$) for the inner, intermediate and outer zone
respectively (for a definition of these zones, see later).

In the above-mentioned papers, it was clearly stated that
the adopted values of $d_{c}$ and $N_{crit}$ had to be
considered not as "solutions" of the problem, but as tentative
guesses to be interpreted in a statistical sense. In particular, to
adopt an identical $d_{c}$ for all the families present in a
given region of the proper element space, could lead to overestimate
the membership of some families, and to underestimate the membership
of some others.

In the present paper, we followed generally the same
procedure, but we have
introduced certain changes to take into account (1) some specificities
of the high-inclination asteroid population; (2) some intrinsic
limits of the statistical approach described above, mainly for what
concerns the definition of the membership of identified families;
(3) the fact that we know a priori that any asteroid sample cannot
be complete beyond some value of magnitude, and family
classifications tend to evolve as larger data sets of asteroid
proper elements, corresponding to increasing number of discovered
objects, become progressively available. In particular:
\begin{itemize}
\item In deriving the QRL, we did not remove \emph{a priori} the members
of big families possibly present in each zone, as was done by
\citet{zappala95}. These authors did so to prevent the possibility
that the presence of very populous families might affect the
generation of the Quasi Random synthetic populations (by
``saturating'' some bins of the proper elements distributions). The
reason why we did not implement this procedure in the present
analysis is that the high-inclination population, mainly in the
inner region, is notably non uniform, and the minimum distance level
achieved by any population of fully-random synthetic objects would
be in any case unreasonably high, leading to possible removal of
very large fractions of the real objects present in some zone.
\item We conservatively adopted as QRL not the average deepest level achieved by
the stalactites of ten quasi-random populations generated in each
zone, but the \emph{deepest} level found in the ten cases.
\item The membership of each family was derived not by looking
simply at the objects present at QRL, but by making a more accurate
analysis of each single case, as explained below.
\item We paid attention to the possible presence of small, but very
compact groupings which might be the cores of families consisting of
small asteroids, most of which have not yet been discovered.
\end{itemize}

As a first step, we divided the main belt into three zones. They
correspond to the inner, middle and outer region of the belt. The
semi-major axis borders between adjacent zones are identical to
those adopted by \citet{zappala95}, and correspond to the 3/1 and
the 5/2 mean motion resonances (MMRs) with Jupiter. The $a_{p}$, $e_{p}$
and $\sin(I_{p})$ boundaries of the three regions are given in
Table~\ref{t:prop}.

Next, for each zone, we defined a corresponding value of
$N_{crit}$. In doing so, we tried to be as consistent as possible
with the values previously used by \citet{zappala95}, taking into
account the differences in both the numbers of objects present in
our regions, as well as the volumes of the regions themselves. In
practical terms, however, we verified that the results of the family
search are not very sensitive to the choice of $N_{crit}$. The
adopted values are listed in Table~\ref{t:prop}.

\begin{table}[h]
\scriptsize
 \centering
  \caption{The properties of the three zones of asteroid belt considered in
this work.} \label{t:prop}
  \begin{tabular}{cccc}
\hline \hline
Parameter & Inner zone& Interm. zone &  Outer zone \\
\hline \hline
 $a_{p_{min}}$~[AU]  & 2.065 & 2.501 &  2.825  \\
 $a_{p_{max}}$~[AU]  & 2.501 & 2.825 &  3.278  \\
 $e_{p_{min}}$       & 0.1   & 0.0   &  0.0    \\
 $e_{p_{max}}$       & 0.35  & 0.35  &  0.4    \\
 $sin(I_{p_{min}})$    & 0.35  & 0.3   &  0.3    \\
 $sin(I_{p_{max}})$    & 0.45  & 0.6   &  0.55   \\
$N_{tot}$            & 3553  & 5439  &  9568   \\
$N_{bins}$           & 8,3,3 & 8,3,3 &  6,3,3  \\
$N_{crit}$           & 12    & 10    &  14     \\
$QRL$~[m/s]          & 130   & 120   &  90     \\
\hline
\end{tabular}
\end{table}

Having now at disposal a value of $N_{crit}$ for each region, we
could derive the corresponding QRL by applying the ``classical''
quasi-random population procedure described above. The resulting 
QRL values, together with the sets of discrete bins in the three 
proper elements adopted in each region, are also listed in Table~\ref{t:prop}.

Once the critical distance level is determined, we have all we need
to perform in each zone our family identification task. We
introduced thus some definitions, and we call \emph{families} the
groups whose stalactites reach at least $QRL-10$~m/s, with a number
of members larger or equal to $N_{crit}$, or reach only QRL, but
having at that level a number of members  equal to at least
$N_{crit}$+2$\sqrt{N_{crit}}$. As an additional requirement, we
impose that a family must in any case be found at QRL, and must be
separated from all other groups existing at that distance level.

We call then \emph{clumps} the groups which marginally fail the
above criteria for family classification. These are groups whose
stalactite reaches QRL, with a number of members larger than
$N_{crit}$ but smaller than $N_{crit}$+2$\sqrt{N_{crit}}$, or groups
which reach $QRL-10$~m/s, with $N_{crit}$ members, but merge with
some other group at QRL. We stress that what we call clumps
are \emph{not} fully flagged families. Many of them are produced by
families which split just below the QRL. In principle, these groups
may be interesting, since it is not clear \textit{a priori} whether
they may simply represent the outcome of the "erosion" of families
or they may correspond to some physical process, including secondary
fragmentation.

The above definition of families, which is strictly related to the
concepts of QRL and $N_{crit}$, is valid in statistical terms only.
In particular, there is a risk that we might miss
some groupings which fail to satisfy the above criteria, but may
well have a physical origin. As an example, very compact, but small
groups, possibly including many objects too small to have been
discovered so far, could produce in stalactite diagrams very deep
stalactite branches, but too thin to satisfy our family definition
requirements. Another case is that of groupings originating within
nominal families as the outcomes of some possible second-generation
disruption event, but being, again, too small to produce stalactite
branches satisfying the above family classification criteria.

The term \textit{cluster} has been used in the past to describe
small and very compact asteroid groupings which are clearly distinct
from the background \citep[see
e.g.][]{PFetal92,Goutelas,zappala94,karin2002}. Accordingly, we
define here as \emph{clusters} the groupings whose corresponding
stalactites may well be noticeably compact and deep to suggest a
possible physical origin, though not formally satisfying our above
definition of families or clumps. We prefer here to be flexible and
we do not introduce a more rigid definition of what we call
clusters. In the next Section, we list for each zone the number of
clusters which we identified from a subjective analysis of our HCM
results. Of course, some arbitrariness can be present, but we think
that in many cases the clusters that we found interesting indeed
deserve further investigation.

We note that the family classification criteria described above are
the result of several experiments performed using slightly different
options. For example, we investigated how different requested
separation between families would influence the results. We found that
the final classification is not very sensitive to small changes in
our requirements. In particular, such changes have very modest
consequences on the resulting list of families in the intermediate
and outer zones. Most families in these zones tend to be quite
robust, and are identified by adopting a large variety of possible
criteria. The situation is somewhat different in the inner zone,
where the list of families may change significantly by using
different classification criteria. This is due to the specific
situation in this region, which is dominated by one large group.

As discussed above, the concept of QRL is quite useful to
identify significant groupings. On the other hand, it has also some
drawbacks as far as the determination of family membership is
concerned. Different families may well be characterized by
differences in the original events that produced them, may have
different ages, may have experienced therefore different evolutions
since their birth, and may be immersed in different environments in
the space of proper elements. As a consequence, in this paper we
have adopted a case-by-case approach, proposed by \citet{nes2005},
to better define the most likely membership of each identified
family.

\begin{figure}[ht]
\centering
\includegraphics[height=10.0cm,angle=-90]{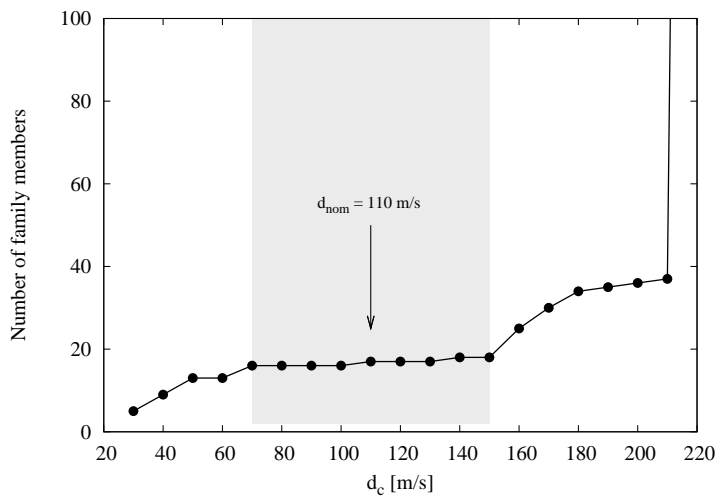}
\caption[]{The number of asteroids belonging to the family around the
asteroid (116763)~$2004EW_{7}$, as a function of the distance cut-off
level $d_{c}$. It can be seen that below $70$~m/s the core of the family
is seen to grow. At $70$~m/s the growth nearly stops and the number of members remains
almost constant until $150$~m/s. At $160$~m/s some limited growth
occurs again, until $210$~m/s when family starts to merge with
the background population in the surrounding region.} \label{f:nfv}
\end{figure}

As an example of this approach, we present here our analysis for the
case of the grouping around the asteroid (116763)~$2004EW_{7}$
located in the intermediate region. The procedure described below is
fully representative of what we did in general, with only a few
exceptions. In Fig.~\ref{f:nfv} we show the resulting number of
members as a function of the adopted distance cut-off level $d_{c}$.
After an initial growth at small distance levels, it is apparent the
presence of a \textit{plateau}, an interval of distance at which the
family membership remains nearly constant, before further growth due
to incorporation of some neighboring group, and a final merging with
the background. The plateau extends from 70 to 150~m/s. In such
cases, we choose as the most appropriate distance level for defining
membership the one corresponding to the center of the plateau. When
two or more plateaus are present (e.g. as in the case of the Tina
family) usually we adopt the value corresponding to the center of
the deepest plateau (the one occurring at smaller distance). As for
the definition of plateau center, whenever the plateau consists of
an even number $n$ of points (we record membership at discrete steps
of $10$~m/s in distance), we conservatively adopted the $n/2$th
smallest distance level in the plateau to define family membership.
Whenever the number $n$ is odd, we simply choose the central
distance value in the plateau.

In a few cases of very complex families (e.g. Euphrosyne) the above
procedure cannot be applied. In these cases we analyzed the
structure of the family, and decided subjectively our preferred
distance level for family membership.

The same method was also used to determine the membership of clumps.
In the case of the clusters, we did not devise any special criterion
to define membership, but we simply looked subjectively at their
stalactite structure, taking profit of the fact that these groups
are very compact over large intervals of distance level.

Before presenting the results of our application of the HCM, let us
note that, as we will show in Section~\ref{s:sdss}, we
have complemented our proper element analysis by taking advantage also
of some additional input, namely the evidence coming from available
SDSS color data for the objects of our sample. This allowed us to
improve our interpretation of proper element data, opened the
possibility to check the consistency of HCM-based family
classification in terms of likely mineralogical composition, and
allowed us to identify in some cases possible random
interlopers.

\section{Summary of HCM results}
\label{s:results}

\begin{figure}[H]
\centering
\includegraphics[height=15.5cm,angle=-90]{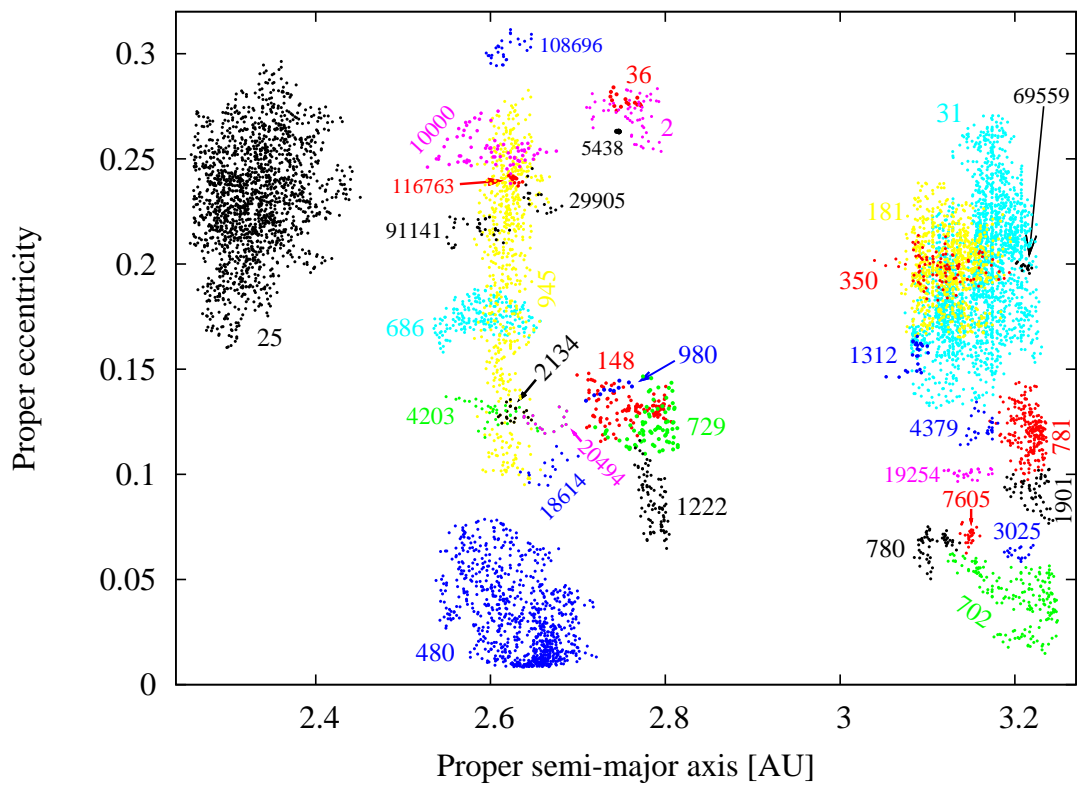}
\caption[]{Locations of the asteroid families, identified in this
work, in the ($a_{p},e_{p}$) plane. Different types and sizes of
symbols are used to distinguish between the members of different
families.}\label{f:fam_ae}
\end{figure}

\begin{figure}[H]
\centering
\includegraphics[height=15.5cm,angle=-90]{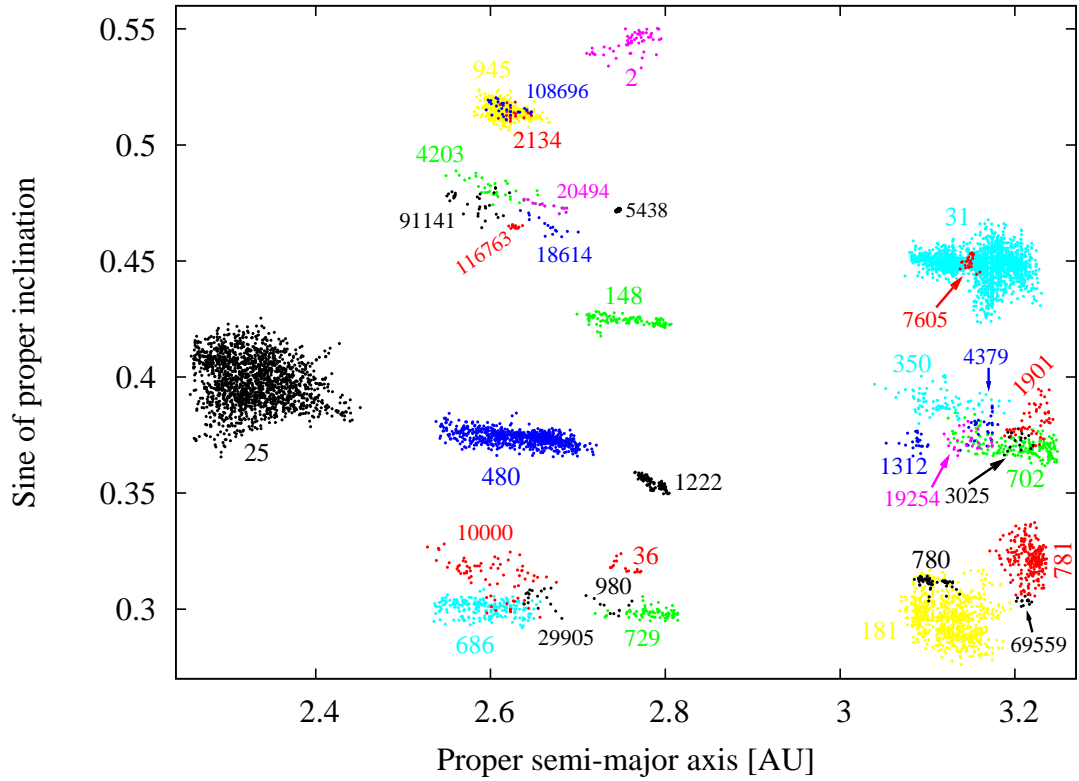}
\caption[]{The same as Figure~\ref{f:fam_ae}, but in the
($a_{p},\sin(I_{p})$) plane.}\label{f:fam_ai}
\end{figure}

In this section we present and discuss the results of our HCM-based
analysis. We split our discussion into three separate sub-sections,
devoted to groupings identified in the inner, intermediate and outer
zone, respectively. The positions of identified asteroid families in the
($a_{p},e_{p}$) and ($a_{p},\sin(I_{p})$) planes are shown in
Figs.~\ref{f:fam_ae} and \ref{f:fam_ai}.

The lists of families, clumps and clusters are
given in Tables~\ref{t:fam},\ref{t:clumps} and \ref{t:clusters},
respectively.\footnote{The complete list of members of all
identified families can be found at:
http://poincare.matf.bg.ac.rs/$\sim$bojan/asteroids/families/high-i-fam/list.html}

\newpage

\begin{center}
\scriptsize
\begin{longtable}{lcccccc}
\caption{List of the identified asteroid families. For each group,
the columns give the lowest-numbered member; the smallest distance
level $d_{min}$ at which it includes $N_{crit}$ objects; the adopted
distance level $d_{nom}$ for membership; the resulting total number
$N$ of members; the proper elements $a_{p}$, $e_{p}$ and
$\sin(I_{p})$ of
the lowest-numbered member.} \label{t:fam} \\
\hline
Name & $d_{min}$& $d_{nom}$& $N$ & $a_{p}$ &  $e_{p}$ &  $\sin(I_{p})$ \\
\hline
\multicolumn{7}{c}{Inner zone} \\
\hline
(25) Phocaea    & 60 & 120 & 1694 & 2.400 & 0.228 & 0.397  \\
(7784) 1994PL   &110 & 120 &   19 & 2.268 & 0.197 & 0.418  \\
\hline
\multicolumn{7}{c}{Intermediate zone} \\
\hline
(2) Pallas            &100 & 120 &  57 & 2.771 & 0.281 & 0.548  \\
(36) Atalante         &120 & 120 &  16 & 2.749 & 0.275 & 0.324  \\
(148) Gallia          & 40 & 120 & 113 & 2.771 & 0.132 & 0.425  \\
(480) Hansa           & 20 & 110 & 839 & 2.644 & 0.009 & 0.375  \\
(686) Gersuind        & 60 & 100 & 207 & 2.589 & 0.173 & 0.302  \\
(729) Watsonia        & 50 & 160 & 139 & 2.760 & 0.123 & 0.299  \\
(945) Barcelona       & 40 & 130 & 600 & 2.637 & 0.251 & 0.512 \\
(980) Anacostia       &110 & 130 &  18 & 2.741 & 0.140 & 0.298  \\
(1222) Tina           & 50 & 120 &  89 & 2.793 & 0.082 & 0.354  \\
(2134) Dennispalm     & 90 & 110 &  19 & 2.638 & 0.130 & 0.512  \\
(4203) Brucato        &100 & 130 &  46 & 2.605 & 0.132 & 0.483  \\
(10000) Myriostos     &110 & 150 &  73 & 2.587 & 0.269 & 0.319  \\
(18614) $1998DN_{2}$  &120 & 120 & 16 & 2.644 & 0.100 & 0.470  \\
(20494) $1999PM_{1}$  &110 & 110 & 14 & 2.684 & 0.127 & 0.473  \\
(29905) $1999HQ_{11}$ &120 & 140 & 28 & 2.675 & 0.226 & 0.299  \\
(89713) $2001YB_{113}$&110 & 110 & 11 & 2.578 & 0.092 & 0.369  \\
(91141) $1998LF_{3}$  &100 & 150 &  30  & 2.599 & 0.222 & 0.474  \\
(108696) $2001OF_{13}$& 90 & 130 &  36 & 2.647 & 0.309 & 0.514  \\
(116763) $2004EW_{7}$ & 50 &  50 &  13  & 2.625 & 0.240 & 0.465 \\
\hline
\multicolumn{7}{c}{Outer zone} \\
\hline
(31) Euphrosyne      & 30 & 100&2063& 3.155 & 0.208 & 0.447  \\
(181) Eucharis       & 50 &  90& 373& 3.128 & 0.217 & 0.305  \\
(350) Ornamenta      & 80 & 110&  93& 3.114 & 0.192 & 0.387  \\
(702) Alauda         & 70 &  90& 179& 3.194 & 0.021 & 0.369 \\
(780) Armenia        & 30 & 120&  76& 3.117 & 0.070 & 0.312  \\
(781) Kartvelia      & 30 &  80& 232& 3.227 & 0.103 & 0.312  \\
(1312) Vassar        & 90 & 100&  24& 3.094 & 0.161 & 0.370  \\
(1444) Pannonia      & 80 & 100&  18& 3.158 & 0.140 & 0.323  \\
(1901) Moravia       & 70 & 80 & 54 & 3.237 & 0.097 & 0.389  \\
(3025) Higson        & 70 &  80&  17& 3.207 & 0.059 & 0.374  \\
(4379) Snelling      & 80 &  90&  29& 3.168 & 0.120 & 0.370  \\
(5931) Zhvanetskij   & 70 &  90&  64& 3.192 & 0.164 & 0.304  \\
(7605) $1995SR_{1}$  & 60 & 120&  30& 3.151 & 0.071 & 0.453  \\
(19254) $1994VD_{7}$ & 80 &  80& 26 & 3.160 & 0.101 & 0.370  \\
(52734) $1998HV_{32}$& 80 &  80& 16 & 3.101 & 0.140 & 0.451  \\
(69559) $1997UG_{5}$ & 60 &  60&  14& 3.214 & 0.198 & 0.304  \\
\hline
\end{longtable}
\end{center} 


\begin{center}
\scriptsize
\begin{longtable}{lcccccc}
\caption{The same as in Table~\ref{t:fam}, but for the identified clumps.
The smallest distance level $d_{min}$ corresponds to $N_{crit}/2$} \label{t:clumps} \\
\hline
Name & $d_{min}$& $d_{nom}$& $N$ & $a_{p}$ &  $e_{p}$ &  $\sin(I_{p})$ \\
\hline
\multicolumn{7}{c}{Inner zone} \\
\hline
(2745) SanMartin  & 90 & 120 & 22 & 2.288 & 0.159 & 0.386  \\
(26142) $1994PL_{1}$&110 & 120 & 13 & 2.264 & 0.176 & 0.385  \\
(100681) $1997YD_{1}$&110 & 110 & 10 & 2.278 & 0.266 & 0.419  \\
\hline
\multicolumn{7}{c}{Intermediate zone} \\
\hline
(194) Prokne            &110 & 120 &  18 & 2.617 & 0.196 & 0.296  \\
(2382) Nonie            & 80 & 110 & 19 & 2.760 & 0.275 & 0.544  \\
(4404) Enirac           & 50 & 110 & 52 & 2.644 & 0.113 & 0.512  \\
(40134) $1998QO_{53}$   & 80 & 150 & 24 & 2.735 & 0.226 & 0.433  \\
(59244) $1999CG_{6}$    &110 & 120 & 11 & 2.634 & 0.165 & 0.471  \\
(62074) $2000RL_{79}$   & 50 & 110 & 33 & 2.586 & 0.091 & 0.372 \\
(81583) $2000HD_{46}$   & 50 & 100 & 44 & 2.616 & 0.162 & 0.512 \\
(103219) $1999YX_{3}$   & 80 & 110 & 13 & 2.642 & 0.082 & 0.371 \\
(114822) $2003ON_{15}$  & 70 & 110 & 24 & 2.740 & 0.139 & 0.424  \\
(195207) $2002DN_{2}$   & 90 & 100 &  5 & 2.565 & 0.114 & 0.479 \\
\hline
\multicolumn{7}{c}{Outer zone} \\
\hline
(1101) Clematis       & 50   & 70 & 16 & 3.242 & 0.034 & 0.369  \\
(1612) Hirose         & 60   & 70 & 20 & 3.102 & 0.115 & 0.307  \\
(2793) Valdaj         & 70   & 80 & 45 & 3.164 & 0.076 & 0.378  \\
(2967) Vladisvyat     & 60   & 80 & 74 & 3.210 & 0.116 & 0.296  \\
(13935) $1989EE$      & 70   & 70 & 10 & 3.140 & 0.261 & 0.450  \\
(14424) Laval         & 80   & 80 & 14 & 3.145 & 0.118 & 0.371  \\
(15161) $2000FQ_{48}$ & 80   &100 & 25 & 3.203 & 0.173 & 0.338   \\
(16243) Rosenbauer    & 90   &100 & 25 & 3.149 & 0.155 & 0.331   \\
(22805) $1999RR_{2}$  & 60   & 70 & 17 & 3.147 & 0.171 & 0.304  \\
(23886) $1998SV_{23}$ & 80   & 80 & 16 & 3.128 & 0.110 & 0.309   \\
(25295) $1998WK_{17}$ & 80   & 90 & 19 & 3.174 & 0.105 & 0.383  \\
(26324) $1998VG_{16}$ & 60   & 90 & 19 & 3.129 & 0.035 & 0.380  \\
(28884) $2000KA_{54}$ & 90   & 90 & 18 & 3.093 & 0.038 & 0.373  \\
(29596) $1998HO_{32}$ & 80   & 80 & 22 & 3.142 & 0.142 & 0.297  \\
(34676) $2000YF_{126}$& 60   & 70 & 15 & 3.211 & 0.157 & 0.297 \\
(35664) $1998QC_{64}$ & 90   & 90 & 14 & 3.112 & 0.063 & 0.371 \\
(38834) $2000SP_{1}$  & 80   & 80 & 16 & 3.125 & 0.199 & 0.386   \\
(52661) $1998BT_{8}$  & 80   & 90 & 18 & 3.109 & 0.072 & 0.373 \\
(55940) $1998GU_{8}$  & 80   & 80 & 27 & 3.170 & 0.157 & 0.435   \\
(58892) $1998HP_{148}$& 60   & 70 & 18 & 3.135 & 0.162 & 0.305 \\
(71193) $1999XG_{231}$& 80   & 90 & 18 & 3.091 & 0.163 & 0.308 \\
\hline
\end{longtable}
\end{center}

\newpage

\begin{center}
\scriptsize
\begin{longtable}{lcccccc}
\caption{The same as in Table~\ref{t:fam}, but for proposed asteroid clusters.
The smallest distance level $d_{min}$ corresponds to $N_{crit}/2$} \label{t:clusters} \\
\hline
Name & $d_{min}$& $d_{nom}$& $N$ & $a_{p}$ &  $e_{p}$ &  $\sin(I_{p})$ \\
\hline
\multicolumn{7}{c}{Inner zone} \\
\hline
(2860) Pasacentennium&100 & 120 & 9 & 2.332 & 0.161 & 0.392  \\
(6246) Komurotoru    &120 & 120 &11 & 2.447 & 0.255 & 0.396  \\
(31359) $1998UA_{28}$& 80 & 100 &11 & 2.272 & 0.200 & 0.403  \\
(58419) $1996BD_{4}$ & 80 & 110 &10 & 2.276 & 0.233 & 0.369  \\
\hline
\multicolumn{7}{c}{Intermediate zone} \\
\hline
(247) Eukrate           & 90 & 110 & 5 & 2.741 & 0.202 & 0.427  \\
(5438) Lorre            & 10 &  50 & 8 & 2.747 & 0.263 & 0.472  \\
(36240) $1999VN_{44}$   & 90 & 110 & 5 & 2.619 & 0.185 & 0.466  \\
(44219) $1998QB_{3}$    & 60 & 110 & 7 & 2.724 & 0.121 & 0.510  \\
(48606) $1995DH$        & 90 & 100 & 5 & 2.668 & 0.109 & 0.478  \\
(76404) $2000FG_{13}$   & 70 &  80 & 6 & 2.623 & 0.191 & 0.299  \\
(91136) $1998KK_{6}$    & 70 & 110 & 6 & 2.613 & 0.145 & 0.483  \\
(103056) $1999XX_{134}$ & 90 & 100 & 9 & 2.623 & 0.281 & 0.512  \\
(109195) $2001QE_{75}$  & 80 &  90 & 7 & 2.656 & 0.084 & 0.373 \\
(208080) $1999VV_{180}$ & 70 &  90 & 6 & 2.608 & 0.119 & 0.513 \\
\hline
\multicolumn{7}{c}{Outer zone} \\
\hline
(24440) $2000FB_{1}$  & 40   & 50 &16 & 3.167 & 0.171 & 0.437  \\
(30575) $2001OM_{101}$& 60   & 60 &11 & 3.124 & 0.045 & 0.380  \\
(59853) $1999RP_{82}$ & 50   &100 &14 & 3.044 & 0.099 & 0.322  \\
(63530) $2001PG_{20}$ & 90   &160 &53 & 2.888 & 0.111 & 0.300  \\
\hline
\end{longtable}
\end{center}

\subsection{Inner zone}

In the inner belt region ($2.065-2.501$~AU) we analyzed a sample of
3553 numbered and multi-opposition asteroids with $\sin(I_{p}) \geq 0.295$.
These objects are separated from low-inclination main belt asteroids belt and from the
neighboring Hungaria region by both mean-motion and secular
resonances \citep{synthpro2,carruba2009}. An inner boundary in semi-major axis, often related to the 7/2
mean motion resonance with Jupiter, is located at about
$2.25$~AU. However, a non-negligible number of asteroids are still
present beyond this limit. The outer boundary, located close to
$2.5$~AU, is set by the powerful 3/1 MMR with Jupiter. Moreover, the
region is delimited by important secular resonances (SRs): the
$\nu_{6}$=$g-g_{6}$ at low inclination, and the $\nu_{5}$=$g-g_{5}$
and $\nu_{16}$=$s-s_{6}$ at high inclination\footnote{The $g$ and
$s$ are the average rates of the longitude of perihelion $\varpi$
and of the longitude of node $\Omega$, respectively. The indexes 5
and 6 refer to planets Jupiter and Saturn, respectively.}
\citep{sec_res1991,michtchenko2010}. Although deep close encounters
with Mars are not possible due to the Kozai class protection mechanism
\citep{milani89}, even shallow encounters may result in removal of
asteroids for eccentricities higher than about $0.3$. These are the
most important dynamical mechanisms which separate the Phocaea group
from the rest of the main belt. Therefore, these high-inclination asteroids seem to
be located in a stability island. Since they are
filling up a bounded stability region and are quite concentrated,
this makes any search for collisional families rather difficult.

Using our procedures, we identified a couple of nominal families,
and a small number of clumps. In addition, we found also several
potentially interesting smaller groupings which are classified as
clusters.

The region is dominated by one single, large group of asteroids,
whose lowest-numbered object is (25) Phocaea. Since this group
contains almost $50\%$ of objects in the region, the identification
of other significant groupings is quite complicated.
This is very similar to the situation that is found in the Hungaria
region \citep{warner2009,milani2010}.

\begin{figure}[h]
\centering
\includegraphics[height=14.5cm,angle=0]{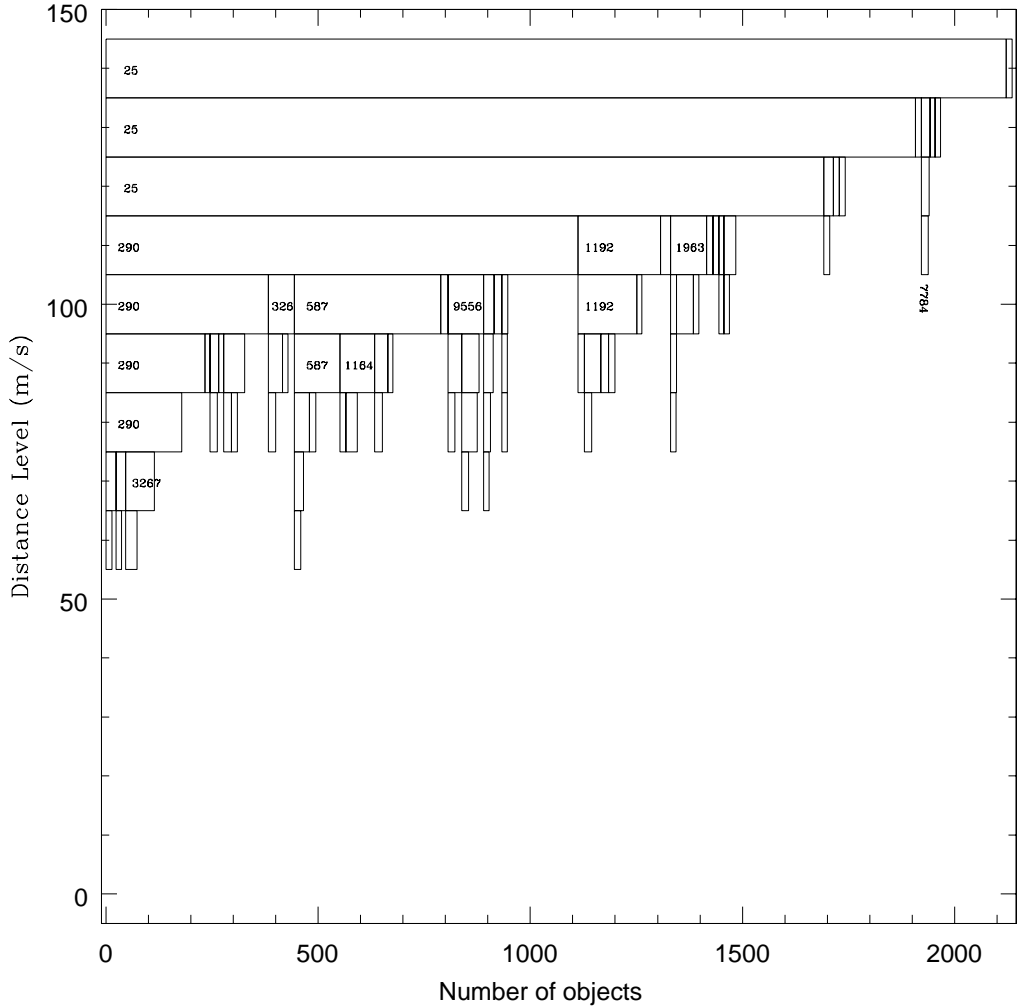}
\caption[]{The stalactite diagram for the inner zone ($N_{crit}$=12,
$QRL = 130$~m/s). At each distance level, only groupings having
at least $N_{crit}$ members are plotted.}\label{f:inner_high}
\end{figure}

In Fig.~\ref{f:inner_high} the resulting stalactite diagram for our
sample is shown. As can be seen, the dominant feature is the large
group of (25) Phocaea. Many minor groups visible in the diagram are
substructures of it, and if Phocaea is a real collisional family, at
least some of its subgroups could represent the outcomes of
second-generation collisions. If this is true, it is likely that
Phocaea is an old family, as was recently suggested by \citet{carruba2009}
who estimated it to be up to $2.2$~Gyr old.

Apart from Phocaea, we identified only one other nominal family
according to our selection criteria. This is a group whose lowest-numbered
object is (7784) 1994PL. This family is identified here for the
first time.

Two families proposed by \citet{gil_hutton2006}, namely Wood and
Krylov are not confirmed. There is a grouping including asteroid
(1660) Wood whose lowest-numbered object is (1192) Prisma, but
this group did not pass our significance criteria.
We did not find any significant grouping associated
with asteroid (5247) Krylov. This asteroid is a member of the
Phocaea family at a distance level of $120$~m/s.

Moreover, a grouping including (2860) Pasacentennium, which was
previously found by \citet{gil_hutton2006}, is fairly small, and we
classify it now tentatively as a cluster, although its stalactite
branch is not very deep. Similarly, a grouping around the
asteroid (6246) Komurotoru classified as a clump by
\citet{carruba2009}, who performed this search in the space of
proper frequencies (see \citet{carruba2007} for details on this
methodology), is included in our list of clusters.
\citet{carruba2009} classified (26142) $1994PL_{1}$ as a clump, and
this is confirmed by our analysis.

None of the other groups proposed by \citet{gil_hutton2006} and
\citet{carruba2009} have passed our significance criteria. Among
these groups there is also (19536) $1999JM_{4}$, classified as a family
by \citet{carruba2009}.

\subsection{Intermediate zone}

In our high-inclination sample, 5439 asteroids belong to the
intermediate zone ($2.501-2.825$~AU). This region is characterized by
a roughly uniform distribution of objects in the semi-major axis
\emph{versus} eccentricity plane, whereas concentrations and gaps are apparent in
the semi-major axis \emph{versus} inclination plane (see
Fig.~\ref{f:aei}). From a dynamical point of view, this zone is
characterized by a mixing of stable and chaotic regions. The eight stable
islands are separated by three mean motion (Jupiter-asteroid) and three linear
secular resonances \citep[see][]{carruba2010b}.

\begin{figure}[h]
\centering
\includegraphics[height=14.5cm,angle=0]{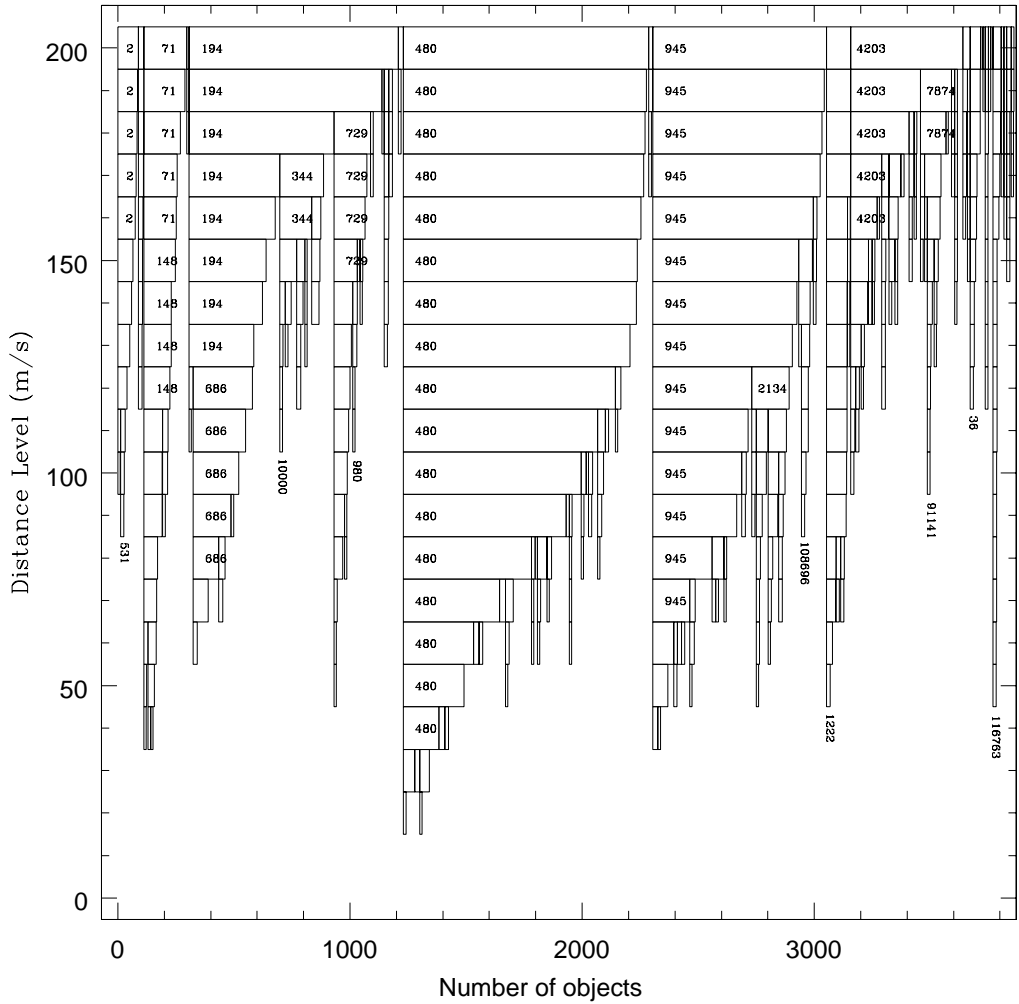}
\caption[]{The stalactite diagram for the intermediate zone
($N_{crit}$=10, $QRL = 120$~m/s). As in Fig.~\ref{f:inner_high},
at each distance level only groupings having at least $N_{crit}$
members are shown.} \label{f:middle_high}
\end{figure}

The stalactite diagram for this zone is shown in Fig.~\ref{f:middle_high}.
We found 19 asteroid families, 10 clumps and 10 clusters.

As can be seen in Fig.~\ref{f:middle_high}, the situation in this
region is very different from what is found in the inner region.
Several important groupings are immediately recognizable. Two large families
are prominent: these are the families of (480) Hansa and (945) Barcelona.
They were already mentioned in the literature. The Hansa
family was originally proposed by \citet{Hergenrother1996}, while a
Barcelona family was first identified by \citet{foglia2004}. Both
families were also confirmed later by other authors
\citep[e.g.][]{gil_hutton2006,carruba2010b}. In addition to Hansa
and Barcelona, however, several sharp and deep stalactite branches
can be seen in Fig.~\ref{f:middle_high}, corresponding to families
whose collisional origin seems very likely.

The (686) Gersuind family, first identified by
\citet{gil_hutton2006} is confirmed, while another group originally
classified as a clump by the same author, Gallia, is now a
full-flagged family.

The families of (1222) Tina and (4203) Brucato, recently proposed by
\citet{carruba2010b}, are confirmed as well.\footnote{In the cases of families
like Tina, whose members interact with one or more 
secular resonances, the proper elements we used are not fully appropriate
and identification results might be different if resonant proper elements \citep{lemaitre1994,carruba_morby2011}
would be used. This, however, seems more important for family membership than for 
recognition of family.} It is
interesting to note that \citet{carruba2010b} identified Brucato
family in the space of proper frequencies only, while in the space
of proper elements he identified it as a clump. We also confirm the
existence of a Watsonia family, mentioned by \citet{cellino2002} on
the basis of spectroscopic properties pointed out by
\citet{Burbineetal92} and \citet{bus1999}. According to still
unpublished observations (Cellino 2011, in preparation) (729)
Watsonia belongs to a rare group of objects, called
\textit{Barbarians} after their prototype, the asteroid (234)
Barbara, which exhibit unusual polarimetric properties
\citep{Cellinoetal06,MasCel09}. Very interestingly, we found in this
region another family, whose lowest-numbered member is (980)
Anacostia, which is also a Barbarian \citep{gil_huttonetal08}.
The Watsonia and Anacostia families merge together well
above the QRL.

In addition, we also found 10 new families, having as their
lowest-numbered objects (36) Atalante,
(2134) Dennispalm, (10000) Myriostos, (18614) $1998DN_{2}$,
(20494) $1999PM_{1}$, (29905) $1999HQ_{11}$, (89713) $2001YB_{113}$,
(91141) $1998LF_{3}$, (108696) $2001OF_{13}$, and (116763) $2004EW_{7}$.

The family proposed by many authors \citep[see
e.g.][]{williams92,lemaitre1994,gil_hutton2006,carruba2010b} around
the very large asteroid (2) Pallas passed our selection criteria as
well. A group including Pallas is present at $QRL-10$~m/s, but it
consists of 8 members, only. However two other groupings associated
to asteroids (531) Zerlina and (1508) Kemi, which have 14 and 23
members at $QRL-10$~m/s respectively, merge with group around Pallas
at QRL forming a group of 57 asteroids.

The Pallas family is certainly interesting in terms of
possible composition, since (2) Pallas belongs to a fairly rare
taxonomic class ($B$). According to \citet{Clarketal10} (2) Pallas
is the largest object belonging to a small number of $B$-class
asteroids which exhibit a blueish trend in the reflectance spectrum
which extends also in the near-IR. No other asteroid which has been
found so far to share this same behavior belongs to our family.
However, we do know that several members of the family are
classified as $B$-class, as pointed out by \citet{Clarketal10}. In
this respect, the member list that we find now is largely in
agreement with that given by the above authors. Spectroscopic
observations extending into the near-IR of members of the Pallas
family will be very interesting to confirm a genetic relationship
with (2) Pallas. Since (2) Pallas is one of the biggest asteroids
(it is actually the biggest one, if (1) Ceres is considered to be a
dwarf-planet) its family could well be another example of the
outcome of an energetic cratering event, as in the well known case
of Vesta. If this is true, it is likely that many members are quite
small and faint (Pallas being a low-albedo object), and have not yet
been discovered. Present and future sky surveys will hopefully be
able to confirm or reject this hypothesis.

In the intermediate zone we have also found numerous clusters. Two
of these, namely (5438) Lorre and (44219) $1998QB_{3}$ are extremely
compact. Both clusters are clearly distinct from any other grouping,
and remain separated even at very large distance levels around
$200$~m/s. These facts suggest a real collisional origin for these
clusters. Moreover, it is known that size and shape of
asteroid families change over time, with respect to the original
post-impact situations. Families slowly spread, and became more and
more dispersed due to the chaotic diffusion and gravitational and
non-gravitational perturbations
\citep{bottke2001,flora2002,carruba03,delloro04}. Being so compact,
it is also likely that the two above-mentioned clusters should be
quite young. We are currently carrying out a detailed study of these
and other clusters, to be presented in a separate paper.

\subsection{Outer zone}
\label{ss:outer}

Our sample includes 9568 high-inclination asteroids in the outer zone
($2.825-3.278$~AU), but most objects are located at
$a_{p} \gtrsim 3.05$~AU. Similarly to the case of the intermediate
zone, the distribution of objects in the outer region is roughly
uniform in the semi-major axis \emph{versus} eccentricity plane,
whereas in the semi-major axis \emph{versus} inclination plane most
asteroids are concentrated in three different dominions. One of them is
located close to $\sin(I_{p}) = 0.3$, another one is centered around
$\sin(I_{p}) = 0.38$, while the third one is centered around
$\sin(I_{p}) = 0.45$ (see Fig.~\ref{f:aei}).

Due to the observed non-uniform distribution in proper inclination, in
our generation of Quasi Random populations in this region we use
only three bins in $\sin(I_{p})$ (see Table~\ref{t:prop}). These
bins have been chosen in such a way that each of them covers one of the
three different dominions in inclination.

\begin{figure}[ht]
\centering
\includegraphics[height=14.5cm,angle=0]{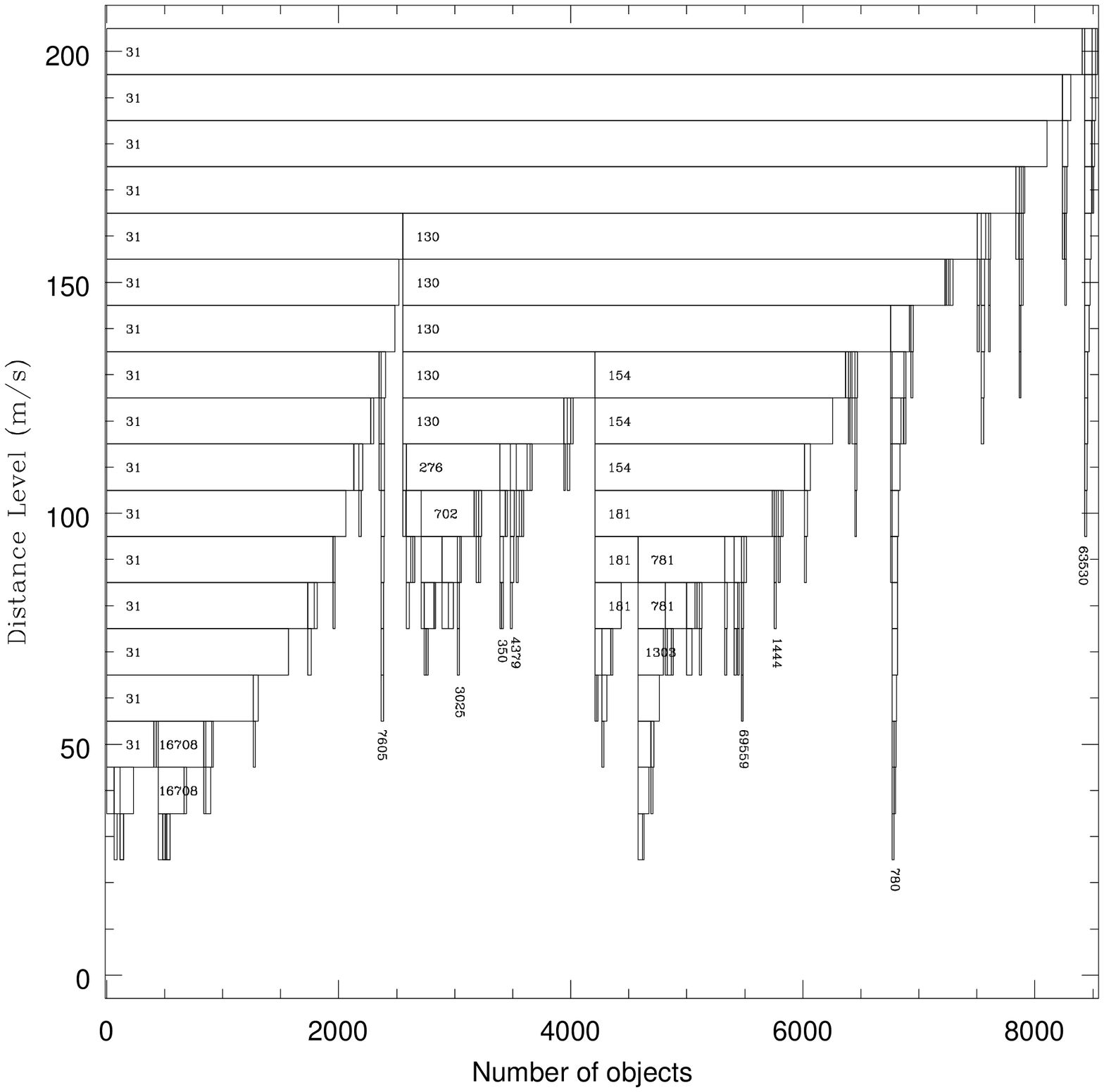}
\caption[]{The stalactite diagram for the outer zone ($N_{crit}$=14,
$QRL = 90$~m/s). As in Figs.~\ref{f:inner_high} and~\ref{f:middle_high},
at each distance level only groupings having at least $N_{crit}$ members are shown.}
\label{f:outer_high}
\end{figure}

The stalactite diagram is shown in Fig.~\ref{f:outer_high}. The
overall structure seems to consist of three major branches merging
together at high distance levels. This might reflect the particular
distribution of the objects in proper inclination. At least
one additional, well separated and very compact group, however, is
also clearly visible as a very deep stalactite branch. The
lowest-numbered member of this family is (780) Armenia.
In total, we identified 17 asteroid families, 21 clumps
and 4 clusters in this region.

The region is dominated by two large families, associated
to asteroids (31) Euphrosyne, and (181) Eucharis.

Among our nominal families, those of Euphrosyne, Alauda and Moravia
had been already identified by other authors
\citep{foglia2004,gil_hutton2006}. It is interesting to note that
asteroid (702) Alauda is known to be a binary system \citep{alauda2007}.
The companion is much smaller than the primary \citep{alauda2010},
suggesting (but this is only a  conjecture) that it might represent
captured ejecta from a collision.

Our families of Eucharis, Pannonia, Filipenko and Snelling
had been classified as clumps in previous investigations
\citep{gil_hutton2006}.

The third largest family in the outer zone is associated to asteroid
(781) Kartvelia. This group is a newly discovered family, but we
note that it merges with Eucharis just $10$~m/s above the critical
QRL. In addition, nine new families have been identified in this
region. These are Ornamenta, Armenia, Vassar, Higson, Zhvanetskij,
$1995SR_{1}$, $1994VD_{7}$, $1998HV_{32}$ and $1997UG_{5}$.

Former asteroid families Weber and (16708) $1995SP_{1}$
\citep{gil_hutton2006}, are now parts of the Hirose clump and the
Euphrosyne family, respectively. We did not find any significant
groupings associated with asteroids (1303) Luthera and
(6051) Anaximenes, which were proposed to be families by
\citet{gil_hutton2006}.

\section{SDSS colors and Spectroscopic data}
\label{s:sdss}

According to previous analyzes available in the literature, it turns
out that, as a general rule, the members of each family tend to
share similar spectral characteristics
\citep[e.g.][]{bus1999,florczak1999,lazzaro1999,ivezic2002,cellino2002}.
Spectral properties of families can thus be used to complement the
results of our HCM analysis of proper elements. In particular,
spectral information may be used both to identify possible family
interlopers as well as to identify objects that might be
candidate family members, although they are not included in nominal
member lists derived by proper element information only
\citep{migliorini1995,milani2010}. Therefore, we have carried out an
analysis of available SDSS colors for the asteroids of our sample.

For the purpose of deriving reliable inferences about asteroid
surface compositions, multi-band photometry is not as precise as
spectroscopy. However, SDSS data are very important, because this
survey includes about two orders of magnitude more objects than
available spectroscopic catalogs. Recently, SDSS data were used by
\citet{roig2006} to identify possible basaltic ($V$-type) asteroids,
and by \citet{parker08} to analyze characteristics of the classical
main belt families. Here, we use the fourth release of the SDSS
Moving Object Catalog~(MOC~4) to analyze the color distribution
properties among different asteroid groups identified in this work.
In cases when spectroscopic data are also available, these are
exploited to reach more reliable conclusions.  In particular, we
used taxonomic/spectral classifications based on SMASS I
\citep{smass1}, SMASS II \citep{taxonomy} and S3OS2 \citep{s3os2}
surveys. In addition, the much older ECAS survey
\citep{tholen1985,tholen1989} was also used whenever possible.

\citet{nes2005} showed that the SDSS MOC is a useful,
self-consistent data-set to study general statistical variations of
colors of asteroids in the main belt, but caution is required to
interpret colors in individual cases. The above authors used an
automatic algorithm of Principal Component Analysis (PCA) to analyze
SDSS photometric data and to sort the objects into different
taxonomic classes.

In particular, PCA can be used to derive linear combinations of the
five SDSS colors, in order to maximize the separation between a
number of different taxonomic classes in SDSS data.\footnote{In
principle, this kind of analysis can also be made using
method adopted by \citet{ivezic2002} and \citet{parker08}. They used
($a^{*}$,$i-z$) instead of ($PC_{1}$, $PC_{2}$) plane, where $a^{*}$
is calculated according to the following relation:
\begin{displaymath}
a^{*} = 0.89(g-r)+0.45(r-i)+0.09(g-i)-0.57\ .
\label{eq:aa}
\end{displaymath}}

According to \citet{nes2005} \citep[see also][]{ivezic2001} the
first two principal components can be used to distinguish among big
taxonomic complexes such as $S$, $C$, or $X$ \citep[see][for
definitions of different taxonomic complexes/classes]{taxonomy}.
These complexes are found to occupy different locations in the
($PC_{1}$, $PC_{2}$) plane.

Following the same procedure, we obtained the relations which define
the two principal components for our sample of asteroids, which
includes 3689 high-inclination objects that are present in the
SDSS MOC~4. The resulting relations are:
\begin{equation}
PC_{1} = -0.337(u-g)+0.470(g-r)+0.618(g-i)+0.533(g-z)\ ,
\label{eq:pc1n}
\end{equation}
\begin{equation}
PC_{2} =-0.654(u-g)+0.489(g-r)-0.305(g-i)-0.491(g-z)\ .
\label{eq:pc2n}
\end{equation}
where $u$,$g$,$r$,$i$,$z$ are the measured fluxes in five SDSS bands
after correction for solar colors; for the values of solar colors
see \citet{ivezic2001}.

\begin{figure}[h]
\centering
\includegraphics[height=12.0cm,angle=-90]{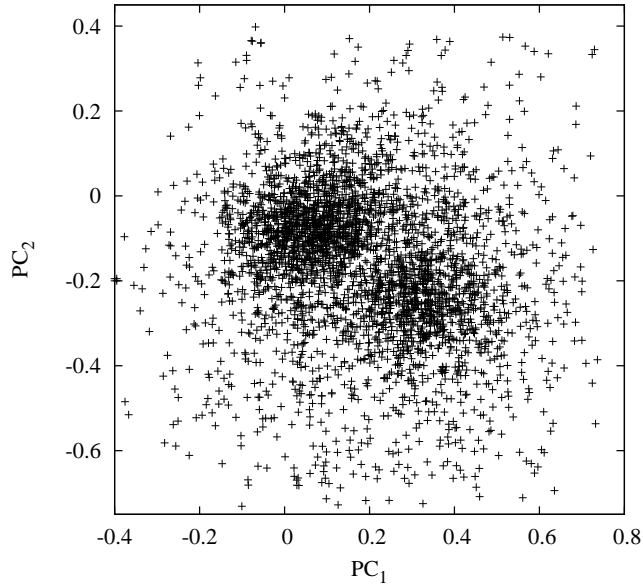}
\caption[]{The distribution of high-inclination asteroids included
in both our sample of proper elements and in the SDSS MOC~4, plotted
in the ($PC_{1}$, $PC_{2}$) plane.}\label{f:sdss4_pc}
\end{figure}

\begin{figure}[ht]
\centering
\includegraphics[height=12.0cm,angle=-90]{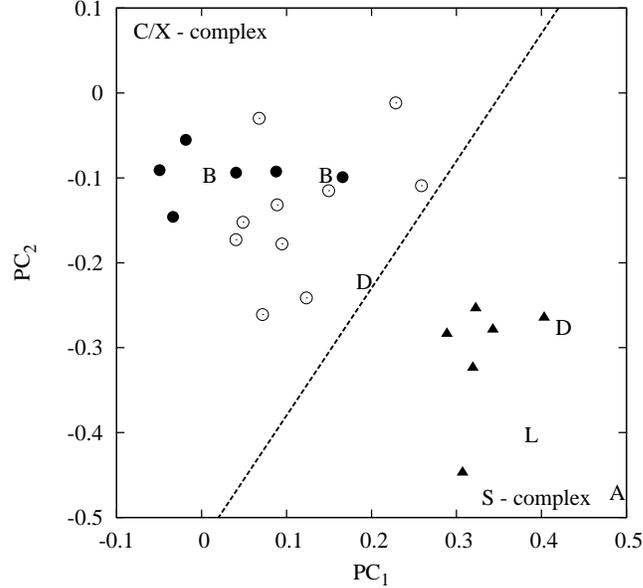}
\caption[]{Locations of the twenty eight high-inclination asteroids
included in the SDSS MOC~4 having known spectral types in the plane
of the principal components obtained according to the
Eqs.~\eqref{eq:pc1n} and~\eqref{eq:pc2n}. Filled circles represent
$C$-type, open circles $X$-type and triangles are $S$-type
asteroids. The locations of two $B$, two $D$, one $A$, and one
$L$-type asteroids are also shown. The separation between $S$ and
$C$/$X$ spectral complexes is quite clear. The approximate border is
shown by the dashed line. Unfortunately, it is not possible to
separate reliably the $C$ and $X$ complexes, because they are mixed
together making it very difficult to find a clear distinction
between them.}\label{f:new_pc}
\end{figure}

In Fig.~\ref{f:sdss4_pc} we plot our sample in the
($PC_{1}$,$PC_{2}$) plane. Two slightly separated, very dense
regions, immersed in a more sparse background, can be easily
recognized. This general behavior was already found by
\citet{nes2005} and \citet{parker08}, who found an association of
the two major groups with different spectral complexes. Following
their example, we plot in Fig.~\ref{f:new_pc} the positions of
asteroids with known spectral types in the plane of our principal
components. From the figure we see that two groups visible in
Fig.~\ref{f:sdss4_pc} correspond to $S$ (bottom right) and $C$/$X$
(top left) complexes.

The $S$ complex appears to be fairly well separated from the $C$/$X$
complex. The situation is worse for the $X$ and $C$ types, which
tend to overlap each other in our PC plane. An $X$ and $C$
overlapping, although slightly less evident, was also found by
\citet{nes2005}.

In Fig.~\ref{f:new_pc} the $C$ complex is typically located at
somewhat smaller values of $PC_{1}$ and higher values of
$PC_{2}$ with respect to $X$ complex objects, but a considerable
mixing of the two complexes is present. As a consequence, we are
generally able to distinguish only among $S$ and $C$/$X$ taxonomic
complexes, although some comments on possible distinctions between
$C$ and $X$ are given in some cases discussed below.

As a first step, we examined the resulting abundance of $S$ and
$C$/$X$ asteroids in the three zones defined in our HCM analysis, in
order to analyze the variation in relative abundance of the main
taxonomic complexes as a function of heliocentric distance. The
results are given in Table~\ref{t:sdss_frac}.

\begin{table}[h]
\scriptsize
 \centering
  \caption{The fraction of objects belonging to
the $S$ and $C$/$X$ taxonomic complexes according to their SDSS
MOC~4 colors, as found in each of three zones, as well as in the
total sample of high-inclination asteroids.} \label{t:sdss_frac}
  \begin{tabular}{ccccc}
\hline \hline
Complex & Inner zone& Interm. zone &  Outer zone & Total \\
\hline \hline
S      & 464 (72\%) & 537 (54\%) &  368 (21\%)  & 1369 (40\%) \\
C/X    & 177 (28\%) & 460 (46\%) & 1408 (79\%)  & 2045 (60\%) \\
Total  & 641        & 997        & 1776         & 3414 \\
\hline
\end{tabular}
\end{table}

As can be seen, and not unexpectedly, the $S$ complex dominates in
the inner zone, whereas the $C$/$X$ complex is dominant in the outer
zone. In the middle region the two complexes have similar
abundances. Although a somewhat larger fraction of $S$ asteroids
could be expected in the intermediate zone, our results are in
general agreement with current knowledge about the abundance of
different taxonomic classes as a function of heliocentric distance
\citep[e.g.][]{taxonomy,md2003}. However, we found a significantly
larger abundance of $S$ asteroids in the outer zone with respect to
recent results by \citet{carvano2010}, who performed a similar study
for low-inclination asteroids in the main belt. According to
\citet{md2003}, differences in the abundance of different taxonomic
complexes across the main belt exist between low- and
high-inclination asteroids. These authors also found that the
abundance of $S$-class asteroids is significantly affected by the
presence of asteroid families. Our results confirm these
findings, although it is not clear to us whether observational
biases acting against the discovery of high-inclination, low-albedo
asteroids in the outer belt could also play an important
role. In any case, we find that a major contribution to the relative
abundance of $S$ class asteroids in the outer belt at high orbital
inclination, is due to the presence of one single, large family,
having as its lowest-numbered member the asteroid (181) Eucharis
(see also the discussion below).

Having the values of principal components, calculated using
Eqs.~\eqref{eq:pc1n} and~\eqref{eq:pc2n}, we could compute
the average values of $PC_1$ and $PC_2$ for all families identified by
HCM for which at least 5 members are included in the SDSS MOC~4.
The corresponding values are listed in Table~\ref{t:sdss}.

\begin{center}
\scriptsize
\begin{longtable}{lccccccc}
\caption{The list of the asteroid families, clumps and clusters, identified in this work,
with available color data in the SDSS MOC~4. Only groups with at
least five members included in the color survey are shown. For each
family, the Table gives: family name; number $N$ of members; number
$N_{SDSS}$ of members observed by SDSS; the values of the principal
components along with their standard deviations; taxonomic complex
according to the values of the SDSS principal components.} \label{t:sdss} \\
\hline
Name & $N$ &  $N_{SDSS}$ & $PC_{1}$ & $\sigma_{PC_{1}}$ &  $PC_{2}$ & $\sigma_{PC_{2}}$ & Taxonomy  \\
\hline
\multicolumn{8}{c}{Inner belt} \\
\hline
(25) Phocaea         & 1694 & 288 & 0.265 & 0.171 & -0.251 & 0.194 & $S$ \\
  \hline
\multicolumn{8}{c}{Intermediate belt} \\
 \hline
(2) Pallas           &  57 & 9  & -0.038 & 0.133 &  0.017 & 0.097 & $C$/$X$  \\
(148) Gallia         & 113 & 22 &  0.290 & 0.165 & -0.330 & 0.175 & $S$  \\
(480) Hansa          & 839 &162 &  0.291 & 0.141 & -0.230 & 0.177 & $S$  \\
(686) Gersuind       & 207 & 40 &  0.390 & 0.118 & -0.279 & 0.145 & $S$  \\
(729) Watsonia       & 139 & 31 &  0.319 & 0.154 & -0.340 & 0.188 & $S$  \\
(945) Barcelona      & 600 & 91 &  0.227 & 0.152 & -0.182 & 0.194 & $C$/$X$ \\
(1222) Tina          &  89 & 17 &  0.120 & 0.217 & -0.130 & 0.171 & $C$/$X$  \\
(4203) Brucato       &  46 & 11 &  0.056 & 0.104 & -0.155 & 0.121 & $C$/$X$  \\
(4404) Enirac        &  52 &  6 &  0.157 & 0.118 & -0.217 & 0.132 & $C$/$X$  \\
(10000) Myriostos    &  73 & 14 &  0.183 & 0.190 & -0.117 & 0.280 & $C$/$X$  \\
(29905) $1999HQ_{11}$&  28 &  9 &  0.255 & 0.201 & -0.249 & 0.161 & $S$  \\
(40134) $1998QO_{53}$&  24 &  6 &  0.200 & 0.255 & -0.293 & 0.149 & $S$  \\
(62074) $2000RL_{79}$&  33 &  9 &  0.306 & 0.080 & -0.183 & 0.150 & $S$  \\
(108696) $2001OF_{13}$& 36 &  5 &  0.167 & 0.087 & -0.273 & 0.147 & $C$/$X$  \\
  \hline
\multicolumn{8}{c}{Outer belt} \\
 \hline
(31) Euphrosyne     &2066& 323 & 0.087 & 0.162  &  -0.045 & 0.174  & $C$/$X$  \\
(181) Eucharis      & 778& 149 & 0.390 & 0.229  &  -0.344 & 0.296  & $S$  \\
(350) Ornamenta     &  93&  14 & 0.025 & 0.179  &  -0.056 & 0.122  & $C$/$X$  \\
(702) Alauda        & 179&  46 & 0.026 & 0.138  &  -0.096 & 0.116  & $C$/$X$  \\
(780) Armenia       &  76&  13 & 0.103 & 0.137  &  -0.063 & 0.170  & $C$/$X$  \\
(781) Kartvelia     & 232&  49 & 0.293 & 0.109  &  -0.150 & 0.150  & $S$  \\
(1101) Clematis     &  16&   5 &-0.025 & 0.056  &  -0.174 & 0.179  & $C$/$X$  \\
(2967) Vladisvyat   &  74&  11 & 0.036 & 0.174  &  -0.120 & 0.230  & $C$/$X$  \\
(3025) Higson       &  17&   5 & 0.062 & 0.106  &  -0.065 & 0.110  & $C$/$X$  \\
(5931) Zhvanetskij  &  64&  20 & 0.058 & 0.166  &  -0.151 & 0.165  & $C$/$X$  \\
(19254) $1994VD_{7}$&  26&   6 & 0.042 & 0.055  &  -0.037 & 0.118  & $C$/$X$  \\
(24440) $2000FB_{1}$&  16&   7 & 0.037 & 0.142  &  -0.079 & 0.082  & $C$/$X$  \\
(25295) $1998WK_{17}$&  19&   6 & 0.145 & 0.234  &  -0.104 & 0.130  & $C$/$X$  \\
(28884) $2000KA_{54}$&  18&   7 & 0.069 & 0.102  &  -0.069 & 0.150  & $C$/$X$  \\
(58892) $1998HP_{148}$&  18&   7 & 0.060 & 0.180  &  -0.214 & 0.220  & $C$/$X$  \\
\hline
\end{longtable}
\end{center}

\begin{figure}[ht]
\centering
\includegraphics[height=12.0cm,angle=-90]{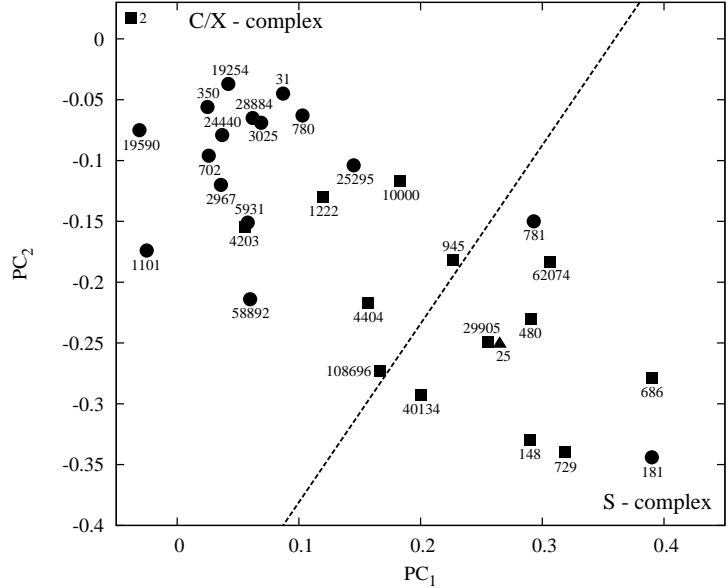}
\caption[]{Locations of the asteroid families, clumps and clusters
identified in this work, in the ($PC_{1}$,$PC_{2}$) plane. The
groups located in the inner, intermediate and outer zone are shown
as triangles, squares and circles, respectively. The dashed inclined
line represents approximately the border between the $S$ and $C$/$X$
complexes.}\label{f:new_pc_fam}
\end{figure}

We found that most families in each zone belong to the dominant
spectral type/complex. This can be better appreciated in
Fig.~\ref{f:new_pc_fam}, in which we show the locations of families,
clumps and clusters in the ($PC_{1}$,$PC_{2}$) plane.

In the inner zone, as expected, all groups belong to the $S$
complex. However, only the Phocaea family has at least five members
included in the SDSS data-set.

More in particular, color data are available for $288$ members of the
Phocaea family. About 80\% of these belong to the $S$-complex. Spectral 
types (derived from spectroscopy) are available for $35$ asteroids that 
belong to the Phocaea family. Most of them are the $S$-class
with only three exceptions, which correspond to likely interlopers.
The asteroids (326) Tamara and (1963) Bezovec are $C$-class, while
asteroid (1318) Nerina is an $X$-class.

In the intermediate zone, the numbers of families belonging to the
$S$ and $C$/$X$ complexes are the same. Seven families
include $S$ members, and seven are $C$/$X$. This is in a good
agreement with the relative abundance of these two complexes in this
region of the belt (see Table~\ref{t:sdss_frac}). The number of
family members with known spectral type 
is very limited. However, in most cases these data are consistent
with derived SDSS colors. Let us discuss here only a few exceptions.

The Gersuind family includes two members with known spectral type,
namely asteroid (686) Gersuind itself, which is $S$ class, and
asteroid (1609) Brenda which is classified as $D$ type. The obtained
values of $PC_{1}$ suggest that members of this family should be
$S$-type asteroids. This is in agreement with the spectral
classification of (686) Gersuind. As for the $D$ classification of
(1609) Brenda, the $D$ class turns out to be a subgroup of the $S$
complex in our Principal Components analysis. Of course, we are
aware that $S$ class asteroids are expected to be quite distinct
from $D$ class objects in terms of thermal history and composition.
$D$ class asteroids have featureless and very reddish spectra, and
are most common among Jupiter Trojans, whereas they are relatively
less common in the main belt. A numerical simulation
performed by \citet{levison} showed that these bodies may have
originated from trans-Neptunian region as a result of the violent
dynamical evolution of the giant-planet orbits as suggested by the
so-called Nice model \citep{menios,morby,gomes}. Interestingly,
results of \citet{levison} suggest an inner boundary for this type
of objects around $2.6$~AU, while asteroid (1609) Brenda has a
semi-major axis of about $2.58$~AU. In any case, we do not rule out
the possibility that either the taxonomic classification of (1609)
Brenda could be wrong, or it may be an interloper in the Gersuind
family.

Reflectance spectra are available also for two members of the
Myriostos family. The asteroid (344) Desiderata turns out to be a
$C$, while (1246) Chaka belongs to the $S$-class. According to SDSS
colors, the members of this family belong to the $C$/$X$ complex, in
agreement with spectral evidence for (344) Desiderata. Therefore, it
is likely that (1246) Chaka is an interloper. Finally, the $C$-class
asteroid (3037) Alku is probably an interloper within the $S$-type
(29905) $1999HQ_{11}$ family.

At least in some cases, very similar values of the principal
components among two groups might suggest a common origin, like, for
example, in the case of Higson family and the (28884) $2000KA_{54}$
clump. They are very close in terms of principal components, and
merge at a distance cut-off of $110$~m/s in the space of proper
elements. This might well be a first example of
spectroscopic confirmation of a genetic relation between a family
and an associated clump. However, available data are not sufficient
to draw definite conclusion in this respect, but we think that new
observations can provide interesting results in the future.

In the outer region almost all identified groups belong to the
$C$/$X$ complex. There are only two exceptions. The Eucharis and
Kartvelia families belong to the $S$ complex. Interestingly, the
Eucharis family is located in a quite peculiar location at the
far edge of the $S$ dominion in the ($PC1$, $PC2$) plane (see
Fig.~\ref{f:new_pc_fam}), ruling out any possibility that it might
have anything to do with the $C$/$X$ complex. Although we cannot
distinguish clearly between the $C$ and $X$ complex in our SDSS
analysis, some indication about a preferred location for the $C$
complex can be drawn. In particular, most families in the outer belt
seem to cluster around a single sub-dominion of the
($PC_{1}$,$PC_{2}$) portion of plane occupied by the $C$/$X$
complexes. Available spectral types are consistent with this
conclusion. The asteroids (350) Ornamenta and (780) Armenia belong
to the $C$ class. Moreover, two members of Euphrosyne and four
members of Alauda that have available spectral types, are all
consistent with the $C$ complex.

The situation of the Eucharis family is in some way unusual. Four
members of this family have known spectral types. Two of them,
including asteroid (181) Eucharis itself, are $X$-class, whereas two
are $C$-class. None of these objects is consistent with an
$S$-class classification inferred for this family from our analysis of
the SDSS colors for the members of this family.

Since it belongs to the $S$ spectral class, which is relatively rare
in the outer belt, the Eucharis family can be clearly distinguished
from nearby background asteroids. According to available SDSS data,
about $25\%$ of the Eucharis family members identified in our analysis
would be interlopers. This is an exceedingly large fraction with
respect to usual situations \citep{migliorini1995,parker08} and
might be an indication of the presence of another separate family
overlapping with Eucharis. Among the suspected interlopers there are
several large asteroids such as (285)~Regina, (746) Marlu,
(1035) Amata, and (29943) $1999JZ_{78}$. The situation is made even
more complicated by the fact that the Eucharis family is located not
far from some well known low-inclination families including
(137) Meliboea and (1400) Tirela. Further investigations, that we
postpone for a future paper, are necessary to address these
questions.

\section{Summary and Conclusions}
\label{s:conclusions}

In this paper a comprehensive search for asteroid families among the
population of high-inclination asteroids has been presented. The
search has been performed by applying the standard Hierarchical
Clustering Method to a sample of 10,265 numbered objects for which
synthetic proper elements were taken from the \textit{AstDys} web
site. To these, we added other 8,295 multi-opposition objects for
which we computed synthetic proper elements. We included in our
sample only asteroids having sine of proper inclination
greater than $0.295$.

We considered three zones corresponding to three different intervals
of proper semi-major axis (inner, intermediate and outer region). We
used the HCM to identify families in each zone. In doing so, we
applied HCM in generally the same way as it was applied in the past
for family searches among the low-inclination population. However,
we also introduced some improvements in the procedure, to achieve a
more reliable and robust classification, mainly for what concerns
family membership. We also make a clear distinction between
highly reliable groupings, that we call families, and more uncertain
ones, that we call clumps. In addition, we call clusters some very
compact groupings for which the number of objects is still low, but
could increase in the future, as more and more objects will be
discovered by observational surveys. The best example of cluster we
found is a very compact eight-members grouping including (5438)
Lorre.

We took advantage of available SDSS MOC~4 color data to improve
family membership reliability and identify likely family interlopers. Using
Principal Component Analysis, we classified all families into $S$ or
$C$/$X$ taxonomic complexes. We found that taxonomical distribution
of families matches very well a systematic variation of asteroid
spectral type with respect to heliocentric distance. Some		
exceptions exist, however, a very interesting case being that of the
Eucharis family.

Asteroid families identified here provide a wide range of opportunities
for possible future studies related to high-inclination asteroids.
Our results are only the first step to fully understanding collisional
evolution of this part of asteroid belt. There is a lot of work
that should be done. For example, to study dynamical characteristics of
proposed families, to estimate their ages, to find size-frequency distributions
(SFDs) of family members, to estimate the size of parent bodies, etc.
These results should be than compared to those obtained for the classical
main belt, what would allow us to understand how differently these
two populations evolved.
Also, typical relative velocity among the high-inclination asteroids
is about $11$~km/s \citep{gil_hutton2006}, while in the classical belt it is
only about $5$~km/s \citep{bottke1994}. Thus, it is interesting to see
how SFDs of high-inclination families fit in numerical experiments
\citep{michel03,durda07}.

Among the individual cases as a particularly interesting to study we
highlight the possibly interplay among Eucharis, Meliboea and Tirela
families. Different chronology methods
\citep{vok2006b,nov2010a,cachucho2010} could be successfully
applied to these groupings.

Some of the studies mentioned above are already possible with existing data,
while some others will be possible in the near future.
Different observational surveys will provide physical characteristics (e.g. albedos,
rotational periods, diameters, spectral types) for many asteroids,
including these on highly inclined orbits. Among these surveys let us mention here
one just finished, Wide-Field Infrared Survey Explorer (WISE), and one planned to
be launched in 2013, Global Astrometric Interferometer for Astrophysics (GAIA).
Other surveys, like this presented recently by \citet{subaru2011}, could
provide valuable data as well.

The results of this investigation open also perspectives for new,
dedicated observing campaigns. In particular, we mention the
interesting case of the Pallas family, which certainly deserves some
further spectroscopic investigations in the visible and near-IR, as
already suggested by \citet{Clarketal10}. Moreover, polarimetric
observations of the Anacostia and Watsonia families in the middle
region, might likely lead to discover new examples of
``Barbarians''. Observations of small compact clusters like Lorre
might be highly welcome as well.

\section*{Acknowledgments}
We would like to thank David Nesvorn\` y and Valerio Carruba,
the referees, for their useful comments and suggestions
which helped to improve this article.
The work of B.N. and Z.K. has been supported by the Ministry
of Education and Science of Serbia, under the Project 176011.

\end{document}